\documentclass[aps,prb,twocolumn,citeautoscript,superscriptaddress]{revtex4}
\usepackage{amsmath}
\usepackage{graphicx}
\usepackage{dcolumn}
\usepackage{float}
\usepackage{color}
\usepackage{multirow}
\usepackage{amssymb}
\usepackage{microtype}
\usepackage{booktabs}
\usepackage[version=4,arrows=pgf]{mhchem}

\begin{document}

\title{Type-II Superconductivity in the Dirac semimetal \ce{PdTe$_2$}}
\author{Ritu Gupta}
\email{ritu.gupta@iitrpr.ac.in}
\affiliation{Laboratory for Muon Spin Spectroscopy, Paul Scherrer Institute, CH-5232 Villigen PSI, Switzerland}
\affiliation{Department of Quantum Matter Physics, University of Geneva, 24 Quai Ernest-Ansermet, 1211 Geneva, Switzerland}
\affiliation{Department of Physics, Indian Institute of Technology Ropar, Rupnagar, Punjab 140001, India}
\author{Catherine Witteveen}
\affiliation{Department of Quantum Matter Physics, University of Geneva, 24 Quai Ernest-Ansermet, 1211 Geneva, Switzerland}
\author{Debarchan Das}
\affiliation{Laboratory for Muon Spin Spectroscopy, Paul Scherrer Institute, CH-5232 Villigen PSI, Switzerland}
\author{Fabian O. von Rohr}
\affiliation{Department of Quantum Matter Physics, University of Geneva, 24 Quai Ernest-Ansermet, 1211 Geneva, Switzerland}
\author{Rustem Khasanov}
\email{rustem.khasanov@psi.ch}
\affiliation{Laboratory for Muon Spin Spectroscopy, Paul Scherrer Institute, CH-5232 Villigen PSI, Switzerland}

\date{\today}

\begin{abstract}
We report on the microscopic superconducting properties of the Dirac semimetal PdTe$_2$. In this study, we have focused on mosaic crystals of PdTe$_2$, and used detailed zero field and transverse field muon spin relaxation/rotation ($\mu$SR), $ac$-magnetic susceptibility, and resistivity measurements to investigate their superconducting properties. The magnetic susceptibility measurements reveal two superconducting transition temperatures at 1.8 and 1.6~K, respectively, in agreement with earlier reports. In contrary to these reports, we find that these mosaic PdTe$_2$ crystals, are not type-I, but rather type-II superconductors. In fact, we observe the clear manifestation of a flux line lattice through a clear diamagnetic shift and Gaussian broadening of the Fourier spectra in the superconducting state. This behavior is likely caused by the disorder in the mosaic crystals of PdTe$_2$ studied here. Our analysis of the superconducting order parameter by the means of temperature dependent magnetic penetration depth $\lambda(T)$ reveals a fully gapped superconducting state that can be well-fitted using an s-wave symmetric gap. We find that PdTe$_2$ is a promising model system for the investigation and interplay of non-trivial topology, surface superconductivity, and type-II bulk superconductivity in a van-der-Waals material. Moreover, our results indicate that the superconductivity in this material can be easily modified from type-I to type-II by disorder in the system.

\end{abstract}

\maketitle
\section{Introduction}

Two-dimensional transition metal dichalcogenides (TMDs) with general formula $MX_2$ ($M$ being a transition metal and $X$ is a chalcogen atom) attracted heaps of attention in the past few years because of their fascinating electronic and optical properties. A vast variety of intriguing, intertwined phenomena such as topological non-trivial electronic band structure, Dirac like dispersion, charge density wave, superconductivity (SC) and many more have been observed in TMDs \cite{yang2017structural,nicholson2021uniaxial,hsu2017topological}. TMDs have been found to especially offer a unique opportunity to explore the unconventional nature of SC. Recent progress in this topic, include the observation of possible chiral SC in 4H$_b$-TaS$_2$ \cite{ribak2020chiral}, the unconventional scaling of the superfluid density with the critical temperature in NbSe$_2$\cite{von2019unconventional}, or the non-trivial topological band structure with SC at 0.1~K in T$_{\rm d}$-MoTe$_2$\cite{guguchia2017signatures}.

In this context, the type-II Dirac semimetal PdTe$_2$ is a material of particular interest as it has been reported to be a potential type-I SC below 1.6 K at ambient pressure\cite{PhysRevB.96.220506}. The type-II Dirac points appear at the contact point of the electron and hole pockets with tilted Dirac cones (tilt parameter $k > 1$ for PdTe$_2$) and are different from the standard type-I Dirac points with point like Fermi surfaces \cite{PhysRevLett.120.156401,bahramy2018ubiquitous}. The possibility of nearly flat bands near the Fermi surface due to tilted Dirac cones and SC below 1.6 K puts PdTe$_2$ as a potential candidate to look for the correlated electronic states, and potentially topological SC. Moreover, evidence of type-II Dirac semimetal in PdTe$_2$ comes via the $ab$ $initio$ electronic structure calculation and angle-resolved photoemission spectroscopy measurements\cite{PhysRevLett.119.016401}.

The type of SC in PdTe$_2$ in presence of a magnetic field is still a matter of debate. The observation of differential paramagnetic effect (DPE) in an applied field (1-$N$)$H_c<H_a<H_c$, where $H_c$ is the thermodynamic critical field and $N$ is the demagnetization factor, provides strong evidence of intermediate state intimating type-I SC in PdTe$_2$\cite{PhysRevB.96.220506}. The signature of an intermediate state has also been observed in recent muon spin relaxation/rotation ($\mu$SR) studies on single crystalline PdTe$_2$\cite{PhysRevB.100.224501}. This is further supported by the low value of Ginzburg-Landau parameter $\kappa=\lambda/\xi$ = 0.09-0.34, where $\lambda$ is magnetic penetration depth and $\xi$ is coherence length\cite{PhysRevB.96.220506,salis2018penetration}. Moreover, coexistence of type-I and type-II SC has been suggested with a range of spatially distributed critical fields by means of scanning tunneling spectroscopy (STS)\cite{sirohi2019mixed,PhysRevB.97.014523} and point-contact spectroscopy\cite{PhysRevB.99.180504}. Observation of a vortex core in another STS experiment point towards the type-II nature of SC\cite{PhysRevLett.120.156401}. 

There are also some evidences in support of type-II/I SC in PdTe$_2$, such as observation of pronounced diamagnetic screening signal in $ac$-susceptibility with applied fields $H_c<H_a$, appearance of SC even after applying fields of the order of 0.3 T that are much higher than the thermodynamic critical field (34.9 mT)\cite{PhysRevB.96.220506}. 
The contradiction in determining the nature of SC through various measurements on different batches of PdTe$_2$ indicates that disorder might play a crucial role in the SC in this material. Indeed, it is well known that the nature of SC is decided through a dimensionless Ginzburg-Landau parameter $\kappa$ which is the ratio of penetration depth $\lambda$ and coherence length $\xi$, i.e., $\kappa = \lambda/\xi$. $\kappa<1/\sqrt{2}$ for type-I and $1/\sqrt{2}<\kappa $ for type-II superconductors. Recently, Tian et al. have shown that reduced mean free path $l$ can result in increased value of $\kappa$ and hence system can be tuned towards type-II SC\cite{PhysRevB.99.180504}. This may provide an excellent opportunity to search for Majorana zero modes, similar to the case of iron-based superconductors, where Majorana zero modes were claimed for surface states with fully gaped SC \cite{PhysRevX.8.041056,zhang2018observation}. Moreover, it has been shown through Pd doping in Pd$_{1-x}$Pt$_x$Te$_2$ that with increasing disorder, type-II SC can be induced in parent type-I PdTe$_2$ \cite{PhysRevB.105.054508}. Therefore, through a careful growth mechanism we have, here, synthesized and characterized mosaic crystals of PdTe$_2$ with low value of residual resistivity ratio which ultimately resulted into reduced mean free path $l$. Hence, the system could be tuned towards type II SC by the reduction of the mean free path. Through $\mu$SR experiments, we were able to determine the characteristic length scale magnetic penetration depth ($\lambda$), the analysis of which reveals a fully-gapped superconducting state in PdTe$_2$

\begin{figure*}[tbh]
\centering
\includegraphics[width=1\textwidth]{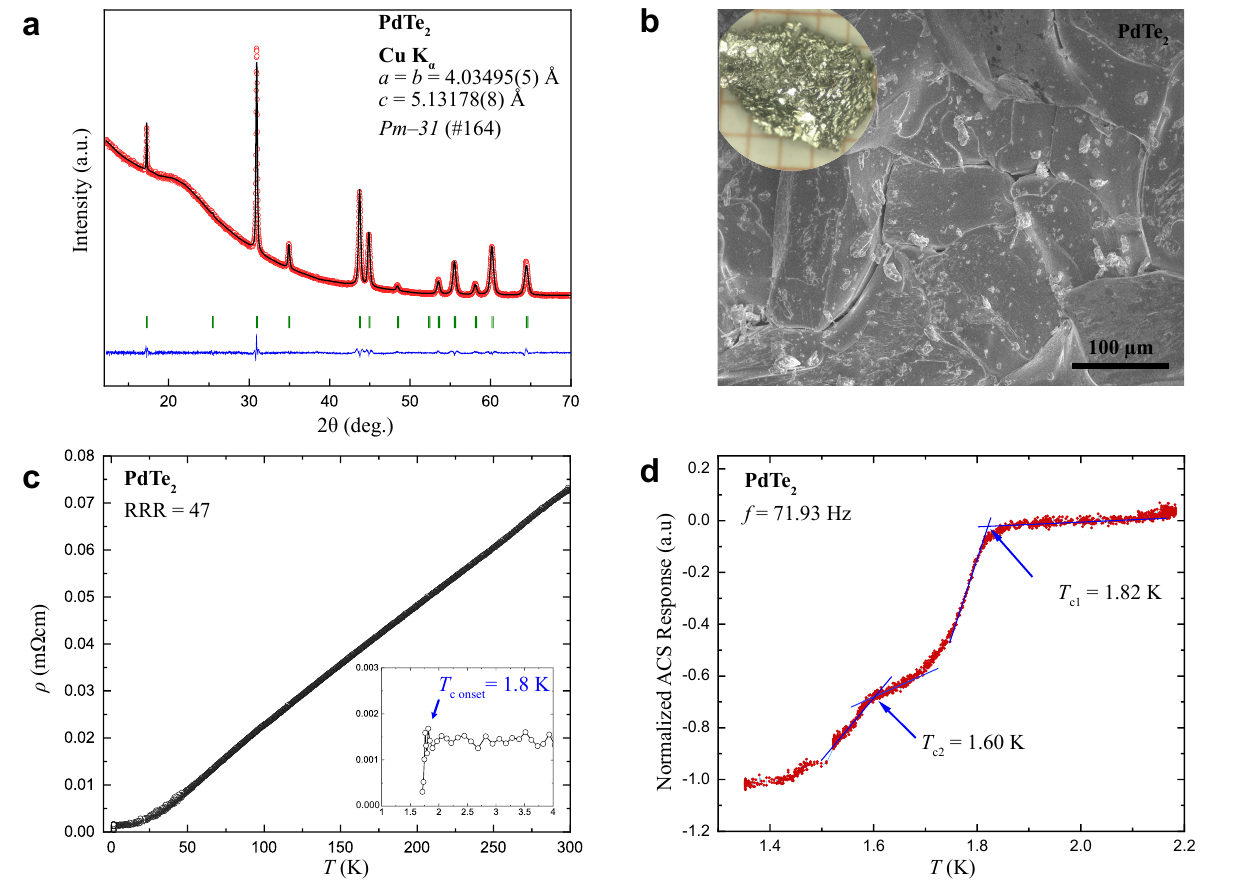}
%\vspace{-1.0cm}
\caption{(a) Powder X-ray pattern of crushed mosaic crystals measured in transmission mode with Cu K\textsubscript{$\alpha$} radiation. (b) Scanning electron microscopy image showing the grain boundaries. Inset: Optical image of the magnification on a mosaic crystal on millimeter paper. (c) Resistivity on a crystal showing the RRR value of 47. Inset: magnified resistivity around the superconducting transition.  (d) Normalized ac susceptibility response as a function of temperature showing two transitions.}%
\label{fig1}
\end{figure*}

\section{Methods section}
The mosaic crystals were prepared using the slow cooling method. Palladium (powder, Merck, 99.995\%) and tellurium (shot, Sigma Aldrich, 99.999\%), 2 g in total, were mixed in a stoichiometric ratio, thoroughly grounded to a homogeneous mixture and sealed under 1/3 atm of argon in a quartz tube (length = 10.5~cm). The sample was then heated with 180 \textdegree C h\textsuperscript{–1} to the temperature of 950 \textdegree C, kept at this temperature for 24h, and cooled at a rate of 5\textdegree C h\textsuperscript{–1} to 600 \textdegree and quenched in air, which resulted in mosaic crystals with metallic luster with dimensions around 6 x 6 x 2 mm\textsuperscript{3} [Fig.~1(b), upper inset].  The phase purity of the samples was checked by means of X-ray diffraction at room temperature on a Panalytical Empyrean diffractometer in capillary mode (Cu K\textsubscript{$\alpha$} radiation and Pixcel linear detector). The chemical composition of the crystals was measured by energy dispersive x-ray spectroscopy (EDX) in a Jeol JSM7600F scanning electron microscope with an EDX detector. The homogeneity of the composition was confirmed over the whole surface of the crystals by multiple point-and-shoot quantitative analyses. The resistance was measured using the four-point probe method on a cryogenic magnet system (Teslatron PT Cryogen Free, Oxford Instrument) in the temperature range between $T$ = 1.7 and 298 K using 10~mA as a current.

The $\mu$SR experiments in zero field (ZF) and transverse field (TF) configurations were carried out at the DOLLY spectrometer at $\pi$E1 beamline, Paul Scherrer Institute, Switzerland. For such experiments, 100 \% spin polarized muon beam with their spin perpendicular to the magnetic filed for TF configuration were used. The ZF-$\mu$SR measurements were done in true zero field maintained by an active compensation system. Few crystals of PdTe$_2$ were mounted on the sample holder and sandwiched between two copper foils to provide good thermal contact to the sample. He-4 cryostat with a He-3 insert was used to lower the temperature down to 270~mK. The DOLLY spectrometer is fully equipped with a standard veto setup, which gives a low background $\mu$SR signal. TF and ZF $\mu$SR spectra were then analyzed using the free software package MUSRFIT\cite{suter2012musrfit}.  

\section{Results and discussion}
\subsection{Mosaic Crystal Characterization}

PdTe$_2$ crystals were synthesized with the slow cooling method as described in the methods section and resulted in mosaic crystals. Powder X-ray diffraction (PXRD) on a powderized crystal confirmed that the correct $CdI_2$ structure type (space group P$\overline{3}$m1, no. 164) was obtained, as well as the phase purity of the samples. The PXRD pattern along with a LeBail fit is presented in Fig. \ref{fig1}(a). No apparent impurities or unreacted starting materials can be detected in the PXRD pattern. The high background towards low angles is a result of the high absorption of the material. In Fig.~\ref{fig1}(b), the micro and macrostructure of the mosaic crystals is shown. In the scanning electron microscopy (SEM) image, the actual surface of the crystals with grain boundaries with a notably short length of approximately 100~$\mu$m is shown. The small grain-sizes of the individual facets are also observable in the optical image shown in the inset of Fig.~\ref{fig1}(b). Upon magnification of an as-grown crystal, the apparent grain boundaries are clearly visible.

The temperature-dependent resistivity of PdTe$_2$ is shown in Fig.~\ref{fig1}(c). The beginning of the resistivity drop is observable at $\approx$ 1.8~K. The residual resistivity ratio here defined as RRR = \textit{R}(298K)/\textit{R}(2K) was determined to be RRR = 47. This value is in contrast to otherwise high RRRs of $>$ 200 for PdTe$_2$ (compare, e.g. reference \cite{PhysRevMaterials.2.114202}). Hence, in our mosaic crystals, the electrons have a much shorter mean free path \textit{l}.

In Fig.~\ref{fig1}(d), the $ac$-susceptibility of the mosaic PdTe$_2$ crystals carried out in a pressure cell with small pressure of 0.12 GPa is depicted. We observe two superconducting transitions at $T_{\rm c} =$ 1.8 and 1.6~K. The former transition temperature is most probably related to the surface SC and the latter is related to the bulk superconducting transition. Similar two-step behavior was also observed in the comprehensive transport and magnetization measurements performed at ambient, as well as under applied pressure \cite{PhysRevB.96.220506,leng2019superconductivity}.
%The indication that the 1.6~K transition is related to surface SC comes from the enhancement of the surface sheath SC for an applied field higher than the critical field $H_{\rm c}$ = 13~mT \cite{PhysRevB.96.220506}. This surface SC does not follow the standard Saint-James–deGennes relation with the critical field, i.e., $H_{\rm c3} = 2.39\times \kappa H\rm c$.   

%\begin{figure}[tbh]
%\centering
%\includegraphics[width=1.0\linewidth]{Fig.1.pdf}
%\vspace{-1.0cm}
%\caption{Normalized ACS response as a function of temperature for close to the ambient value of pressure. Inset shows the picture of a sample holder used for $\mu$SR experiments at DOLLY spectrometer with single crystals of PdTe$_2$ mounted.}%
%\label{fig:Sample-PressureCell}%
%\end{figure}

\begin{figure*}[tbh]
%\centering
\includegraphics[width=1\linewidth]{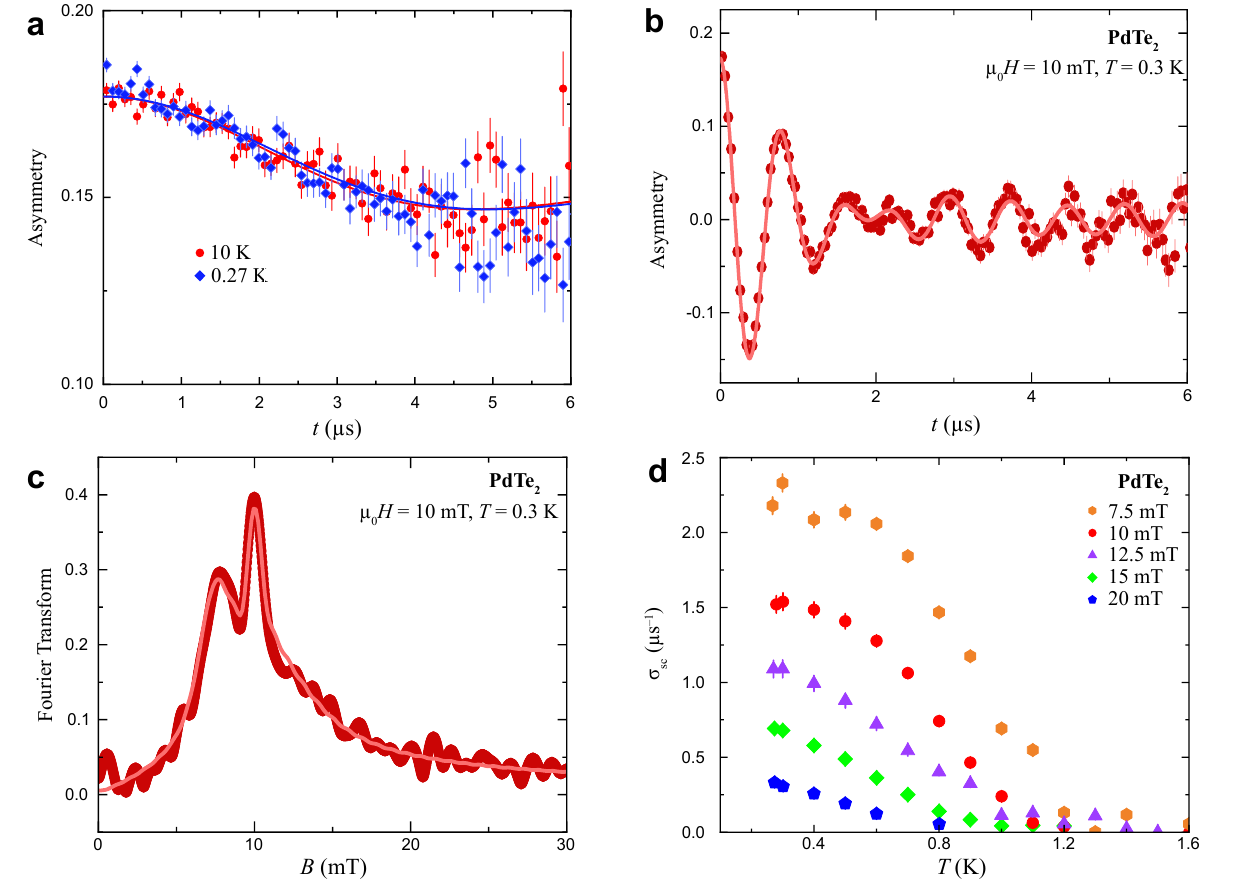}
%\vspace{-1.0cm}
    \caption{ (a) Asymmetry-time $\mu$SR spectra in zero applied field collected much above (10 K) and below (0.27~K) the superconducting transition. Solid lines through the data points are the fits using Eq.~1. (b) Asymmetry-time $\mu$SR spectra collected on single crystals of PdTe$_2$ with an applied magnetic field $\mu_0H$ = 10~mT in the superconducting state. (c) Fast Fourier transform signal of the corresponding $\mu$SR spectra shown in (b). The solid lines through the data points in (b) and (c) are the fits using Eq.~3. (d) Temperature-dependent Gaussian relaxation rate associated to the superconducting state under various mentioned magnetic fields.  }%
\label{fig:Sample-PressureCell}%
\end{figure*}

\subsection{ZF $\mu$SR experiments}
$\mu$SR experiments in true zero field were carried out in order to look for spontaneous magnetic fields which might emerge due to time reversal symmetry breaking in the superconducting ground state. Fig.~2(a) presents two representative spectra collected above $T_{\rm c}$, i.e. at $T$ = 10~K and inside the superconducting state. i.e., at $T$ = 0.27~K. The spectra were fitted using a Gaussian-Kubo-Toyabe function times the Lorentzian function:
\begin{equation}
P\textsuperscript{ZF}(t)= G\textsubscript{KT}\textrm{exp}(-\lambda t).
\end{equation}

This equation reflects the muon-spin depolarization rate associated with the nuclear magnetic moments, with the following functional form:

\begin{equation}
G\textsubscript{KT}(t) = \frac{1}{3}+\frac{2}{3}(1-\sigma\textsubscript{ZF}^2t^2)\text{exp}\bigg(-\frac{\sigma\textsubscript{ZF}^2t^2}{2}\bigg).
\end{equation}

Here $\sigma\textsubscript{ZF}$ represents the width of the nuclear dipolar field distribution experienced by the muon-spin ensemble. The Lorentzian term accounts for the electronic relaxation in the system. It can be seen that two spectra fall on top of each other with no additional relaxation in the superconducting state. This observation states that time-reversal symmetry is preserved in the Dirac semimetal PdTe$_2$ across the superconducting transition.

%\begin{figure}[tbh]
%\centering
%\includegraphics[width=1.0\linewidth]{ZF.pdf}
%\vspace{-1.0cm}
%\caption{a.) Asymmetry-time $\mu$SR spectra in zero applied field collected much above (10 K) and below (0.27 K) the superconducting transition. Solid line through %the data points are the fits using Eq. 1.}%
%\label{fig:Sample-PressureCell}%
%\end{figure}
\begin{figure*}[tbh]
%\centering
\includegraphics[width=1.0\linewidth]{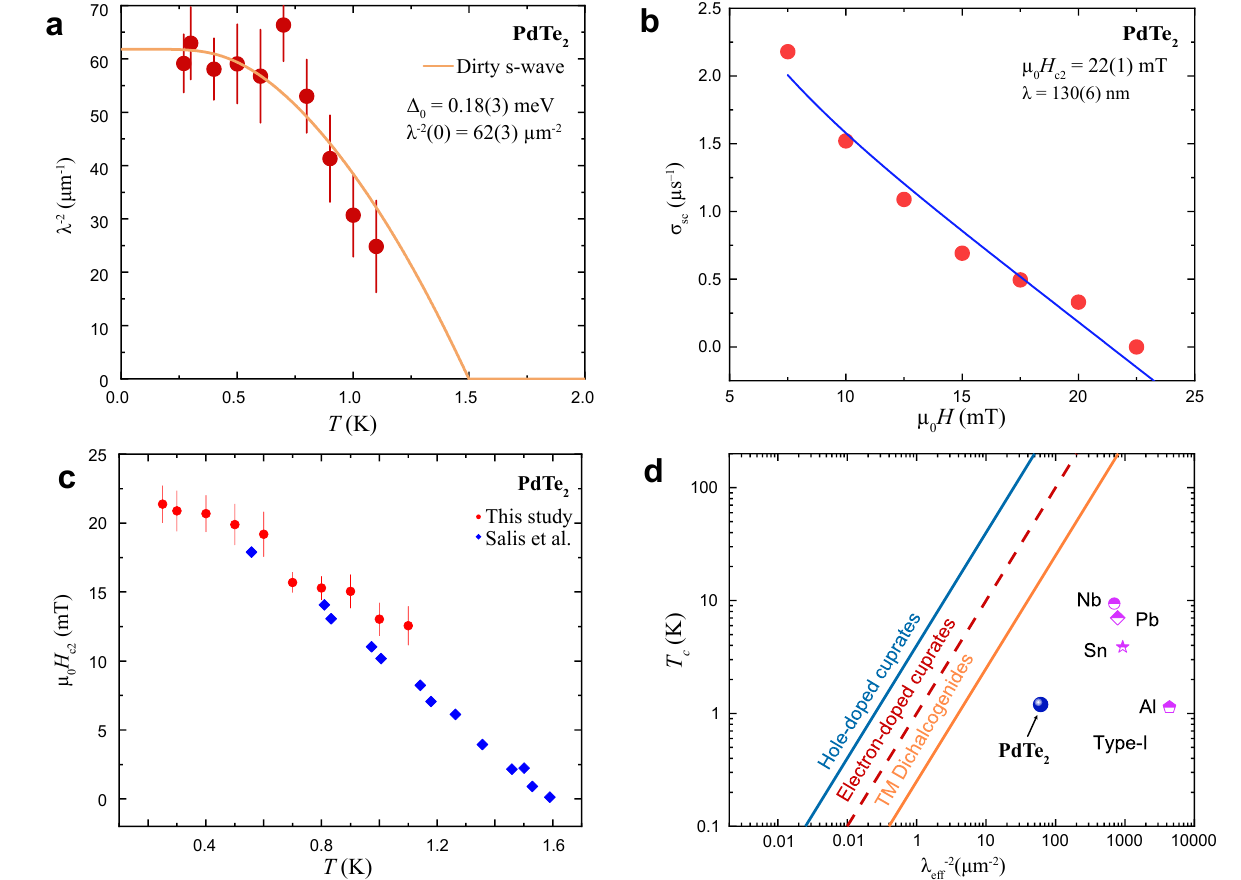}
%\vspace{-1.0cm}
\caption{(a) Temperature dependent inverse squared magnetic penetration depth $\lambda^{-2}(T)$ with a $s$-wave gap fitting using $\alpha$-model in dirty limit. Various parameters obtained after fitting are mentioned in the legend. (b) Magnetic field dependent superconducting Gaussian relaxation rate $\sigma_{\rm sc}(H)$ fitted with Eq.~6. (c) Upper critical field as a function of temperature compared with that obtained by Salis et al. on a disordered Pd$_{1-x}$Pt$_x$Te$_2$ sample.\cite{PhysRevB.105.054508} (d) Uemura plot showing $T_{\rm c}$ vs $\lambda^{-2}_{\rm eff}$ for cuprates, TMDs, and elemental superconductors. The solid-blue, dashed red, and orange lines display the $T_c/\lambda^{-2}_{\rm eff}$ ratio respectively for hole-doped cuprates, electron-doped cuprates\cite{PhysRevLett.101.097009}, and TM dichalcogenides \cite{guguchia2017signatures} with values 4, 1, and 0.4 K $\mu$m$^2$. The large solid blue sphere represents the data for PdTe$_2$.}%
\label{fig:Sample-PressureCell}%
\end{figure*}
\subsection{TF $\mu$SR experiments}

We next focus on the TF $\mu$SR experiment in the superconducting state of PdTe$_2$. Fig.~2(b) shows the TF $\mu$SR spectra accumulated in the superconducting state with an applied field of $\mu_0 H$ = 10~mT perpendicular to the initial spin polarization of the muons. Fig.~2(c) shows the fast Fourier transform (FFT) of the asymmetry-time spectra shown in Fig.~2(b). The FFT spectra show typical behavior expected in case of type-II SC rather than type-I SC, where the latter in the intermediate state must show three peaks: a peak at zero field related to part of the sample being in the Meissner state, at applied field related to background (if any), and at the critical field for the part of the sample in the normal state\cite{PhysRevResearch.2.023142,PhysRevB.104.L100508}.  
The observation of type-II nature of SC in our samples grown using a solid-state method with slow cooling approach is different from what was observed in case of the crystals grown using modified Bridgman technique\cite{PhysRevB.100.224501}. It is well established that different growth techniques induce various type of disorder in the system. Nucleation is typically reduced in Bridgman growth processes as only a limited section of the liquid is initially subjected to the temperature where nucleation can happen. This approach leads to a substantial mass of single domain which differs from the growth method we employed. In our crystals, the slow cooling method and subsequent quenching at 600 \textdegree C results in PdTe$_2$ with very slight variation in stoichiometry, leading to a disordered phase. This can be seen from the Pd-Te phase diagram, where the single phase region is still large at this temperature. These variations in stoichiometry are leading to many defects that are not visible neither in PXRD, nor in EDX analysis, but impact the mean free path and the superconducting properties of the material. We have identified intrinsic disorder, characterized by multiple domains, each typically sized at 100 $\mu$m [Fig.~1(b)]. The $\mu$SR spectra obtained on single crystals grown by Leng $et$ $al.$\cite{PhysRevB.100.224501} were analysed employing a three-component muon depolarization function mentioned above. This observation, characteristic of the intermediate phase in a type-I superconductor, provides compelling evidence for type-I behaviour in the bulk of the PdTe$_2$ crystal with minimal intrinsic disorder. The type-II nature of SC is also in contrast with another $\mu$SR studies, where the coexistence of type-I and type-II SC was observed. The reason for this coexistence is the inhomogeneous mean free path $l$ on the surface of the PdTe$_2$ crystals \cite{singh2019coexistence}. Thus, observation of type-II SC in our study is most convincingly because of the low value of the electron mean free path introduced due to the intrinsic disorder in the system. Additionaly, using the Drude model formula, $l = \frac{3\pi h}{2\rho e^2 k_F^2}$, we estimated a mean free path of 267 nm, which is less than the lower bound of 341 nm predicted by Tian et al.\cite{PhysRevB.99.180504} for a type-I superconductor. Thus, the observed type-II behavior in our crystals aligns with expectations.

Coming back to the TF $\mu$SR experiments, homogeneous formation of vortex lattice was achieved by collecting the spectra during warming up after cooling the sample in the desired field applied above $T_{\rm c}$. Spectra collected above $T_{\rm c}$ (not shown here) shows weak relaxation because of the randomly distributed field from nuclear magnetic moments. The enhancement in the relaxation in the superconducting state is observed because of the inhomogeneous distribution of field created by the flux line lattice (FLL) in the Shubnikov phase. The observed asymmetry-time spectra are well-fitted using a multi-Gaussian fit with three components:
%%%%%%%%%%%%%%%%%%%%%%%%%%%%%%%%%%%%%%%%%%%%%%%%%%%%%%%%%%%%%%%
\begin{equation}
A_{TF}(t)= \sum\limits_{i=0}^3 A_i~e^{-\frac{\sigma_{i}^2t^2}{2}}\cos(\gamma_{\mu}B_it+\varphi),
\label{A1}
\end{equation}
%%%%%%%%%%%%%%%%%%%%%%%%%%%%%%%%%%%%%%%%%%%%%%%%%%%%%%%%%%%%%%%%
where $A_i$, $\sigma_i$, and $B_i$ are respectively the asymmetry, the relaxation rate and the mean field of the $i$th component. $\varphi$ is the initial phase of the muon-spin ensemble and $\gamma_{\mu}/(2\pi)$ = 135.5342 MHz/T is the muon gyromagnetic ratio. The mean field and second moment of such field distribution are given by:
%%%%%%%%%%%%%%%%%%%%%%%%%%%%%%%%%%%%%%%%%%%%%%%%%%%%%%%%%%%%%%%
\begin{equation}
\langle B \rangle= \sum\limits_{i=0}^3 \frac{A_i~B_i}{A_1+A_2+A_3},
\label{eq2}
\end{equation}
%%%%%%%%%%%%%%%%%%%%%%%%%%%%%%%%%%%%%%%%%%%%%%%%%%%%%%%%%%%%%%%%
%%%%%%%%%%%%%%%%%%%%%%%%%%%%%%%%%%%%%%%%%%%%%%%%%%%%%%%%%%%%%%%
\begin{equation}
\langle \Delta B \rangle^2= \frac{\sigma^2}{\gamma_{\mu}^2}=\sum\limits_{i=0}^3 \frac{A_i}{A_1+A_2+A_3}[(\sigma_i/\gamma_\mu)^2+ (B_i-\langle B\rangle)^2].
\label{eq3}
\end{equation}
%%%%%%%%%%%%%%%%%%%%%%%%%%%%%%%%%%%%%%%%%%%%%%%%%%%%%%%%%%%%%%%%
The relaxation rate associated with the superconducting state ($\sigma_{\rm sc}$) is estimated by subtracting a temperature-independent nuclear magnetic moment contribution ($\sigma_{\rm nm}$) using $\sigma_{\rm sc}$ =$\sqrt{\sigma^2-\sigma_{\rm nm}^2}$. Temperature dependence of $\sigma_{\rm sc}(T)$ for magnetic fields 7.5, 10, 12.5, 15, and 20~mT are shown in Fig.~2(d). Inverse squared magnetic penetration depth $\lambda^{-2}(T)$ is then extracted by fitting the field dependent $\sigma_{\rm sc}(H)$ at each temperature using the following equation developed by Brandt\cite{PhysRevB.68.054506}:
\begin{equation}
\sigma_{\rm sc}(\mu s^{-1})\approx 4.83\times 10^4(1-h)\times[1+1.21(1-\sqrt{h})^3]\lambda^{-2}(nm^{-2}),
\end{equation}
where $h=H/H_{\rm c2}$ is the reduced critical field. The temperature dependent $\lambda^{-2}(T)$ obtained by such a procedure is shown in Fig.~3(a). We have noted two superconducting transitions at temperatures of 1.8 K and 1.6 K in the ac susceptibility measurements as a function of temperature [Fig. 1(d)]. Interestingly, in $\mu$SR measurements, a bulk technique, we did not detect any signature of a higher transition. This suggests that the transition occurring below 1.8 K is likely associated with surface superconductivity. 

The superconducting gap symmetry is further elucidated by analyzing $\lambda^{-2}(T)$ data using an isotropic $s$-wave case in the dirty limit\cite{PhysRevB.103.144516}: 

\begin{equation}
\frac{\lambda^{-2}(T)}{\lambda^{-2}(0)}=\frac{\Delta(T)}{\Delta(0)}\tanh\bigg[\frac{\Delta(T)}{k_B T}\bigg],
\end{equation}

where the temperature dependence of gap function is \cite{PhysRevB.102.014514, CARRINGTON2003205, PhysRevB.102.144515, PhysRevB.103.174511,PhysRevB.97.184509,gupta2022microscopic} $\Delta(T)=\Delta(0)\tanh\{1.785[(T_{\rm c}/T-1)^{0.51}]\}$. As temperature dependence of magnetic penetration depth shows clear tendency of saturation at low temperature, we ruled out the possibility of $d$-wave scenario. The temperature-dependence of $\lambda^{-2}(T)$ can be better described by a nodeless $s-$wave model in dirty limit with obtained parameters, shown in the upper inset of the Fig.~3(a). The gap to $T_{\rm c}$ ratio $\Delta_0/\text k_\textrm B T_{\rm c}$ = 1.6, which is smaller than the Bardeen–Cooper–Schrieffer expected value of 1.764 and is in fair agreement with that reported previously through scanning tunneling spectroscopy \cite{PhysRevB.97.014523}. The fully gaped symmetry with weak electron phonon coupling suggests a conventional pairing mechanism in PdTe$_2$. 

Furthermore, the field dependence of TF-relaxation rate $\sigma_{\rm sc}(H)$ collected at 0.27~K [shown in Fig.~3(b)] could be fitted well with the Eq.~6 developed by Brandt for superconductors possessing single gap $s$-wave symmetry. From the fitting, the obtained value of upper critical field ($\mu_0H_{\rm c2}$) and effective penetration depth ($\lambda_{\rm eff}$) at lowest measured temperature (0.27~K) are 22(1)~mT and 130(6)~nm, respectively. Ginzburg-Landau coherence length ($\xi$) is estimated around 115~nm using the formula $\xi = \big(\frac{\phi_0}{2\pi \mu_0 H_{\rm ¢2}}\big)^{1/2}$. The value of Ginzburg-Landau parameter $\kappa = \lambda/\xi$ = 1.13, which is a signature of type II SC in PdTe$_2$. Fig.~3(c) presents the temperature variation of upper critical field $\mu_0H_{\rm c2}(T)$ obtained by fitting Eq.~6 to the $\sigma_{sc}(T,H)$ data shown in Fig.~2(d). For comparison, $\mu_0H_{\rm c2}(T)$ is also presented for a disordered sample showing type-II SC\cite{PhysRevB.105.054508}. 

Relatively low value of $T_{\rm c}$ and high value of the superfluid density $n_s$ ($n_s \propto \lambda_{\rm eff}^{-2}$) are further indicative of conventional nature of SC in PdTe$_2$. From the obtained value of $\lambda_{\rm eff}^{-2}$, we can estimate the superfluid density $n_s$ from the relation $\lambda_{\rm eff}^{-2}=m^*/(\mu_0n_se^2)$, where $m^*$ is the effective mass of the electrons. Effective mass can be estimated via using relation $m^*=(1+\lambda_{\rm e-ph})m_e$, where $m_e$ is the rest mass of the electron and $\lambda_{\rm e-ph}$ is the electron-phonon coupling constant found to be 0.58 from recent helium atom scattering measurements on PdTe$_2$ where $\lambda_{\rm e-ph}$ was approximated from the Debye-Waller attenuation of the He specular peak\cite{npj}. We obtained a value of $n_s$ = 2$\times$10$^{27}$m$^{-3}$ which is very close to other 2D transition materials such as 1.67$\times$10$^{26}$m$^{-3}$ for T$_d$-MoTe$_2$, 2.8$\times$10$^{26}$m$^{-3}$ for 2M-WS$_2$, 1.41$\times$10$^{27}$m$^{-3}$ for $\alpha$-PdBi$_2$\cite{PhysRevLett.127.217002}. Hall measurements will be useful to estimate the normal state charge carriers density which will ultimately allow one to compare the carrier density in normal and superconducting state. This will also be useful in determining the type of charge carriers responsible for the formation of SC in PdTe$_2$, which is a topic of debate until now.    

Moreover, in Fig.~3(d), we present $T_{\rm c}$ vs $\lambda_{\rm eff}^{-2}$ in so called Uemura plot\cite{PhysRevLett.68.2712,PhysRevB.38.909} for PdTe$_2$ along with other well-known superconductors such as hole and electron doped cuprates, transition metal dichalcogenides, and elemental superconductors. It can be seen that $T_c/\lambda_{\rm eff}^{-2}$ (K/$\mu$m$^{-2}$)$\approx$ 0.0195 for PdTe$_2$ lies pretty close to that observed for the elemental superconductors and falls in the range of standard BCS superconductors where the predicted value of $T_c/\lambda_{\rm eff}^{-2}$ (K/$\mu$m$^{-2}$) should be 0.00025–0.015. All these observations together suggest type-II SC with conventional nature in PdTe$_2$. 

\section{Conclusion}
Microscopic superconducting properties of the Dirac semimetal PdTe$_2$ have been investigated by means of ac-susceptibility, resistivity, as well as ZF and TF $\mu$SR experiments carried out on mosaic crystals. In contrary to earlier reports, we found clear evidence, i.e., the formation of a flux line lattice and the resulting Shubnikov phase for type-II superconductivity with a bulk superconducting transition of $T_{\rm c}$ = 1.6~K. The absence of any additional ZF $\mu$SR relaxation in the superconducting state compare to its normal state value confirms that time reversal symmetry is preserved. A fully-gapped superconducting state is observed from our TF $\mu$SR experiment, which can be well-described in a dirty s-wave scenario. Interestingly, the low value of Uemura ratio $T_c/\lambda_{\rm eff}^{-2}\approx$ 0.0195 is pretty close to most of the elemental superconductors. Furthermore, the field dependence of Gaussian relaxation $\sigma_{\rm sc}(H)$ shows the typical behavior observed in case of single gap $s$-wave superconductor with ideal triangular vortex lattice. Thus, our results jointly classify PdTe$_2$ as a time reversal invariant with conventional fully gap superconductor. Our results show that PdTe$_2$ is a promising model material for the investigation and interplay of type-I and type-II superconductivity with topologically non-trivial electronic bands in a layered van-der-Waals material.

\section{acknowledgements}
This work was performed at Swiss Muon Source (S$\mu$S), Paul Scherrer Institute (PSI, Switzerland). The research was supported by the Swiss National Science Foundation through the grants SNF-Grant No. 200021-175935 and PCEFP2\_194183. R.G. would like to thank the financial support from IIT Ropar. We want to thank Prof. Radovan Cerny for helpful discussions and Dr. Adrien Waelchli for his valuable help with the resistivity measurement on the Teslatron.
%\bibliography{citations}

\end{document}